\documentstyle[eqsecnum,aps,twocolumn,epsf]{revtex}
\draft
\begin{document}
\twocolumn[\hsize\textwidth\columnwidth\hsize
\csname @twocolumnfalse\endcsname
\preprint{HEP/123-qed}
\title{Crystal growth, transport properties and crystal structure of the single-crystal La$_{2-x}$Ba$_x$CuO$_4$ with $x=0.11$}
\author{T. Adachi, T. Noji, and Y. Koike}
\address{Department of Applied Physics, Graduate School of Engineering, Tohoku University,\\Aoba-yama 08, Aoba-ku, Sendai 980-8579, Japan}
\date{\today}
\maketitle
\begin{abstract}
We have attempted the crystal growth by the traveling-solvent floating-zone (TSFZ) method of La$_{2-x}$Ba$_x$CuO$_4$ with $x\sim1/8$, where the superconductivity is strongly suppressed. 
Under flowing O$_2$-gas of a high pressure (4 bars), we have succeeded in growing single crystals of $x=0.11$ with 5 mm in diameter and 20 mm in length. 
It has been found that both in-plane and out-of-plane electrical resistivities of the {\it single-crystal} La$_{2-x}$Ba$_x$CuO$_4$ with $x=0.11$ exhibit a clear jump at $\sim$ 53 K. 
The temperature corresponds to the structural phase transition temperature between the orthorhombic mid-temperature (OMT) and tetragonal low-temperature (TLT) phases, $T_{d2}$. 
It has also been found that both in-plane thermoelectric power and Hall coefficient drop rapidly at $T_{d2}$ and decrease below $T_{d2}$ with decreasing temperature. 
These results are quite similar to those observed in the single-crystal La$_{1.6-x}$Nd$_{0.4}$Sr$_x$CuO$_4$ with $x\sim1/8$, suggesting that the so-called static stripe order of holes and spins in the CuO$_2$ plane is formed below $T_{d2}$ in La$_{2-x}$Ba$_x$CuO$_4$ with $x\sim1/8$ as well as in La$_{1.6-x}$Nd$_{0.4}$Sr$_x$CuO$_4$ with $x\sim1/8$.
\end{abstract}

\vspace*{1pt}
\vspace{25pt}
]

\section{Introduction}

Since the discovery of the anomalous suppression of superconductivity in La$_{2-x}$Ba$_x$CuO$_4$ with $x\sim1/8$, \cite{mooden,kumagai} the so-called 1/8 anomaly has been a subject of considerable research attention. 
In recent years, the 1/8 anomaly has been found not only in a series of La-214 high-$T_c$ superconductors \cite{crawford,koike} but also in the Bi-2212 \cite{akoshima,watanabe} and Y-123 ones, \cite{tallon,akoshima2,akoshima3} which means that the 1/8 anomaly is common to high-$T_c$ superconductors including the CuO$_2$ plane in their crystal structures. 
In La$_{2-x}$Ba$_x$CuO$_4$ with $x\sim1/8$, the structural phase transition from the orthorhombic mid-temperature (OMT, space group: Bmab) to tetragonal low-temperature (TLT, space group: P4$_2$/ncm) phase occurs at $\sim$ 70 K. \cite{axe,suzuki,suzuki2} 
Moreover, at low temperatures below the structural phase transition temperature, $T_{d2}$, the transport properties exhibit various anomalous behaviors, such as an increase in the electrical resistivity and decreases in the thermoelectric power and the Hall coefficient with decreasing temperature. \cite{sera} 
Therefore, the intimate connection between the crystal structure, the electronic state and the suppression of superconductivity has attracted notice. \cite{koike2,koike3} 
Several years ago, a static stripe order of holes and spins in the CuO$_2$ plane was discovered from the elastic neutron-scattering experiment in La$_{1.6-x}$Nd$_{0.4}$Sr$_x$CuO$_4$ with $x=0.12$. \cite{nature,prb}
The static stripe order is formed as the TLT phase appears. 
Therefore, they have proposed that the dynamical stripe correlations, observed in a wide range of hole concentration in the La-214 superconductors, \cite{cheong,mason,thurston,yamada} are pinned by the TLT structure, leading to the appearance of the static stripe order and the suppression of superconductivity. 

In La$_{2-x}$Ba$_x$CuO$_4$ with $x\sim1/8$, such a static stripe order has not yet been found in the neutron scattering experiment because of the difficulty in the preparation of a large-sized single crystal of good quality. 
However, it is thought that the static stripe order exists also in this system, taking account of the appearance of the TLT phase and transport anomalies below $T_{d2}$ in the polycrystalline sample which are analogous to those observed in La$_{1.6-x}$Nd$_{0.4}$Sr$_x$CuO$_4$. \cite{crawford,nakamura,koike4} 
Moreover, the observation of a magnetic order of Cu spins in the muon-spin-rotation ($\mu$SR) measurements provides a strong circumstantial evidence in favor of spin stripe ordering in La$_{2-x}$Ba$_x$CuO$_4$ with $x\sim1/8$. \cite{kumagai2,luke} 
Another interest in La$_{2-x}$Ba$_x$CuO$_4$ is that the effect of the Nd spin in La$_{1.6-x}$Nd$_{0.4}$Sr$_x$CuO$_4$ on the formation of the static stripe order can be examined. 
It is because Sakita {\it et al.} have pointed out that the behavior of the susceptibility of La$_{1.6-x}$Nd$_{0.4}$Sr$_x$CuO$_4$ is quite different from that of La$_{2-x}$Sr$_x$CuO$_4$ even above $T_{d2}$, which suggests that the Nd moment may have a great influence on the formation of the static stripe order. \cite{sakita}
Accordingly, the preparation of the high-quality single crystals of La$_{2-x}$Ba$_x$CuO$_4$ with $x\sim1/8$ and the detailed study on the possible stripe order are of much importance. 

Several attempts to grow single crystals of La$_{2-x}$Ba$_x$CuO$_4$ by the traveling-solvent floating-zone (TSFZ) method have been reported so far. \cite{yu,ito,khan,tanabe} 
However, no single crystal of good quality with $x\sim1/8$ has been grown in flowing O$_2$-gas of low pressures below $\sim$ 2 bars. 
Therefore, we have attempted to grow single crystals of La$_{2-x}$Ba$_x$CuO$_4$ with $x\sim1/8$ by the TSFZ method in flowing O$_2$-gas of high pressures, on the analogy of the successful growth of La$_{2-x}$Sr$_x$CuO$_4$ single crystals. \cite{tanaka,tanaka2,kojima,hosoya,kawamata} 
Then, we have investigated the transport properties of the obtained single crystals. 

\section{Experimental details}\label{sec2}

In the crystal growth of La$_{2-x}$Ba$_x$CuO$_4$ with $x\sim1/8$, La$_2$O$_3$, BaCO$_3$ and CuO powders were used as raw materials of the feed rod and the solvent. 
For the feed rod, the powders in the molar ratio of La : Ba : Cu = 1.875 : 0.125 : 1 were mixed and prefired in air at 900 $^{\circ}$C for 12 h.
After pulverization, the prefired materials were mixed and sintered in air at 1100 $^{\circ}$C for 24 h.
This process of mixing and sintering was repeated 4 times to obtain homogeneous powders of La$_{2-x}$Ba$_x$CuO$_4$.
Next, 1 mol \% CuO powders were added to the powders of La$_{2-x}$Ba$_x$CuO$_4$ and mixed thoroughly, in order to obtain tightly sintered feed rods in the sintering process and also to compensate for evaporated CuO in the TSFZ growth process.
The obtained fine powders were put into thin-wall rubber tubes and formed into cylindrical rods under a hydrostatic pressure of 2.4 kbars.
The typical dimensions of the rods were 6 mm in diameter and 120 mm in length. 
In the growth process of La$_{2-x}$Ba$_x$CuO$_4$, one of the most serious problems is deep penetration of the molten zone into the feed rod, which makes the molten zone unstable. 
To avoid such a situation, it is important for the feed rod to be sintered as tightly as possible. 
Therefore, we measured the melting temperature of La$_{2-x}$Ba$_x$CuO$_4$ with $x=0.125$, and the final sintering was carried out at 1250 $^{\circ}$C just below the melting temperature for 24 h in air.

For the solvent rod, the composition of the raw materials was richer in Cu; typically La$_{1.875}$Ba$_{0.125}$ : Cu = 3 : 7 in the molar ratio. 
Powders of the raw materials were mixed and prefired in air at 900 $^{\circ}$C for 12 h. 
After pulverization, the prefired materials were mixed and formed into cylindrical rods. 
Then, the final sintering was performed in air at 900 $^{\circ}$C for 12 h. 
The sintered rods were sliced in pieces and a piece of $\sim 0.4$ g was used as a solvent for the TSFZ growth.

The TSFZ growth was carried out in an infrared heating furnace equipped with a quartet ellipsoidal mirror (Crystal Systems Inc., Model FZ-T-4000-H). 
Under flowing O$_2$-gas of a high pressure (4 bars), the zone traveling rate was 1.0 mm/h and the rotation speed of the feed rod and the grown crystal was 30 rpm in the opposite direction.

In order to fill up oxygen vacancies and to remove the strain, the as-grown crystals were post-annealed in flowing O$_2$-gas of 1 bar at 900 $^{\circ}$C for 50 h, cooled down to 500 $^{\circ}$C at a rate of 8 $^{\circ}$C/h, kept at 500 $^{\circ}$C for 50 h and then cooled down to room temperature at a rate of 8 $^{\circ}$C/h.

The dc magnetic susceptibility was measured with a superconducting quantum interference device (SQUID) magnetometer (Quantum Design, Model MPMS-XL5). 
Electrical resistivity measurements were carried out by the dc four-probe method. 
The thermoelectric power was measured by the dc method with a temperature gradient of $\sim 0.5$ K across a crystal. 
The Hall coefficient was measured by the ac method with a frequency of 30 Hz. 
Powder x-ray diffraction measurements were also performed in a temperature range between 10 K and 280 K, in order to estimate the structural phase transition temperatures. 

\section{Experimental results}
\subsection{Characterization of grown crystals}

We succeeded in keeping the molten zone stable under flowing O$_2$-gas of 4 bars during the TSFZ growth. 
An as-grown crystal is shown in Fig. 1. 
The dimensions were 5 mm in diameter and 70 mm in length.  
A few days later, however, the crystal of a half or over was broken into pieces in air, because the initially grown portion of the crystal contained some inclusions such as La$_2$O$_3$, which was confirmed by the powder x-ray diffraction measurements. \cite{la2o3}
A single crystal of 5 mm in diameter and 20 mm in length was obtained from the part grown in the last stage.
\begin{figure}[htbp]
\begin{center}
\mbox{\epsfxsize=0.45 \textwidth \epsfbox{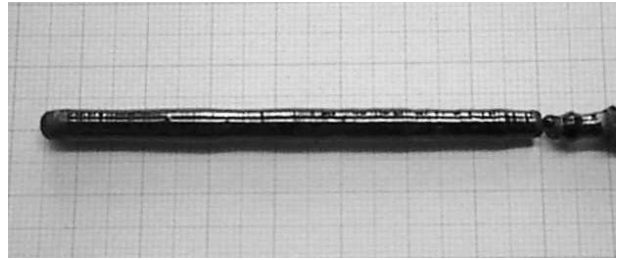}}
\end{center}
\caption{As-grown crystal of La$_{2-x}$Ba$_x$CuO$_4$ by the TSFZ method. The size of the smallest scale is 1 mm.}  
\label{fig:fig1} 
\end{figure}

Details of the domain structure of the grown crystal were investigated by the x-ray back-Laue photography.
As shown in Fig. 2, four-fold symmetric spots were clearly observed in the photograph of a surface parallel to the growth direction, indicative of the crystal as a single crystal. 
Moreover, the broadness of the spots was similar to that in La$_{2-x}$Sr$_x$CuO$_4$ single crystals. \cite{kawamata} 
We also checked the crystal by the powder x-ray diffraction. 
There could be seen Bragg peaks of the La-214 phase and no impurities such as La$_{1+x}$Ba$_{2-x}$Cu$_3$O$_{7-\delta}$, La$_4$BaCu$_5$O$_{13}$ and La$_2$O$_3$ reported in some earlier papers. \cite{yu,ito,tanabe} 
Accordingly, it is concluded that the quality of the grown single crystal is good. 

The Ba content of the grown crystal was estimated by the inductively-coupled-plasma atomic-emission-spectrometry (ICP-AES) to be $x=0.11$, which is a little smaller than that of the feed rod. 
It may be due to some evaporation of Ba and also due to concentration of Ba into the molten zone in the TSFZ growth process. 
In fact, the Ba content in the molten zone was estimated by the ICP-AES after the growth to be La : Ba = 0.94 : 1.06 in the molar ratio. 
Therefore, it is considered that the analyzed ratio of the solvent of La : Ba = 0.94 : 1.06 is more appropriate than the starting composition of La : Ba = 1.875 : 0.125 for the crystal growth of $x=0.11$. 
This content of Ba is much larger than the Sr content in the molten zone in the case of the growth of the overdoped La$_{2-x}$Sr$_x$Cu$_{1-y}$Zn$_y$O$_4$; typically La : Sr = 1.5 : 0.5 in the molar ratio. \cite{kawamata}\begin{figure}[htbp]
\begin{center}
\mbox{\epsfxsize=0.35 \textwidth \epsfbox{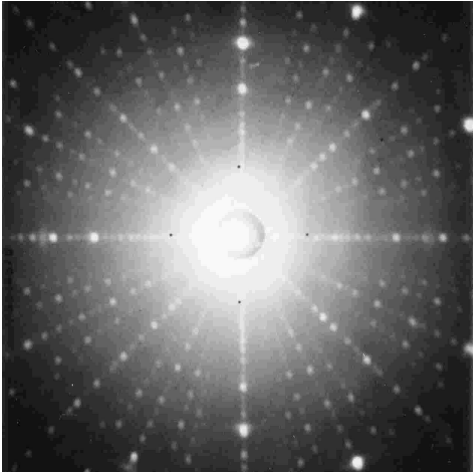}}
\end{center}
\caption{X-ray back-Laue photograph of a surface parallel to the growth direction of La$_{2-x}$Ba$_x$CuO$_4$.}
\label{fig:fig2} 
\end{figure}

The Cu content in the molten zone after the growth was estimated by the ICP-AES to be La$_{0.94}$Ba$_{1.06}$ : Cu = 2.0 : 8.0 in the molar ratio, which is richer than that in the starting composition. 
It has been reported, on the other hand, that the starting composition of the solvent (La, Ba) : Cu = 2 : 8 results in the inclusion of (La, Ba)$_2$Cu$_2$O$_5$ in the grown crystals. \cite{watauchi} 
Therefore, the starting composition of the solvent (La, Ba) : Cu = 3 : 7 may be quite suitable for the crystal growth of $x=0.11$. 

The oxygen content of the post-annealed crystal of La$_{2-x}$Ba$_x$CuO$_4$ was checked by iodometry. 
As a result, there was no oxygen deficiency within the experimental accuracy.
That is, $\delta = 0.00 \pm 0.01$ for La$_{2-x}$Ba$_x$CuO$_{4-\delta}$ with $x = 0.11$, indicating that post-annealing process mentioned in Sec. \ref{sec2} is enough to obtain stoichiometric single crystals of La$_{2-x}$Ba$_x$CuO$_4$. 

Figure 3 shows the temperature dependence of the zero-field-cooled (ZFC) and field-cooled (FC) dc magnetic susceptibilities of La$_{2-x}$Ba$_x$CuO$_4$ with $x=0.11$ in a magnetic field of 10 Oe. 
It is found that superconductivity of the bulk appears in this sample. 
The superconducting transition temperature, $T_c$, defined as the cross point between the extrapolated line of the steepest part of the ZFC superconducting transition curve and zero susceptibility, is 10.2 K.
\begin{figure}[htbp]
\begin{center}
\mbox{\epsfxsize=0.4 \textwidth \epsfbox{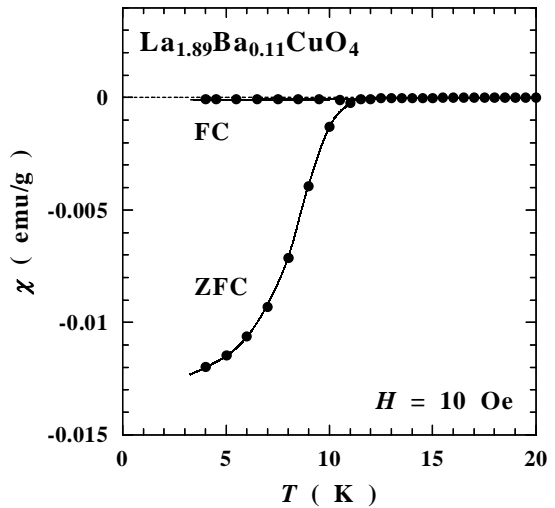}}
\end{center}
\caption{Temperature dependence of the zero-field-cooled (ZFC) and field-cooled (FC) dc magnetic susceptibilities of La$_{2-x}$Ba$_x$CuO$_4$ with $x=0.11$. The solid lines guide the reader's eye.}
\label{fig:fig3} 
\end{figure}

\subsection{Transport properties and crystal structure}

Figure 4 shows the temperature dependence of the in-plane ($\rho_{ab}$) and out-of-plane ($\rho_c$) electrical resistivities in the single-crystal La$_{2-x}$Ba$_x$CuO$_4$ with $x=0.11$. 
The $\rho_{ab}$ exhibits a metallic behavior in the normal state. 
The residual resistivity, defined as the extrapolated value of the resistivity in the normal state to $T$ = 0 K, is as small as $\sim10^{-5}$ $\Omega$cm, which is similar to those of high-quality single crystals of La$_{2-x}$Sr$_x$CuO$_4$. \cite{kawamata,kimura} 
The $\rho_c$ exhibits a semiconducting behavior in the normal state. 
A kink is observed at 256 K, corresponding to the structural phase transition temperature between the tetragonal high-temperature (THT, space group: I4/mmm) and OMT phases, $T_{d1}$. \cite{kambe} 
The anisotropy, $\rho_c/\rho_{ab}$, is $\sim$ 10$^3$ at room temperature, which is almost the same as that of La$_{2-x}$Sr$_x$CuO$_4$ with $x\sim1/8$. \cite{kimura} 
The value of $T_c$, defined as the midpoint of the superconducting transition curve, is 17.8 K. \cite{tc} 
The values of $T_c$ and $T_{d1}$ are in good agreement with those of the polycrystalline sample with $x=0.11$, respectively, as shown in Fig. 5. \cite{axe,koike5} 
A clear jump is observed at $\sim$ 53 K in the temperature dependence of both $\rho_{ab}$ and $\rho_c$, as shown in the inset of Fig. 4, though no jump has been observed in the polycrystalline samples with $x\sim1/8$. 
The jumps are considered to be due to the structural phase transition between the OMT and TLT phases, as in the case of the single-crystal La$_{1.6-x}$Nd$_{0.4}$Sr$_x$CuO$_4$ with $x=0.12$. \cite{nakamura} 
In fact, the temperature is roughly in correspondence to $T_{d2}$ of the polycrystalline sample with $x=0.11$ as also shown in Fig. 5. \cite{axe,billinge} 
These results on the electrical resistivity also suggest that the composition of the single crystal is almost the same as that of the polycrystalline sample with $x=0.11$ and that the quality of the single crystal is quite good. 
\begin{figure}[htbp]
\begin{center}
\mbox{\epsfxsize=0.4 \textwidth \epsfbox{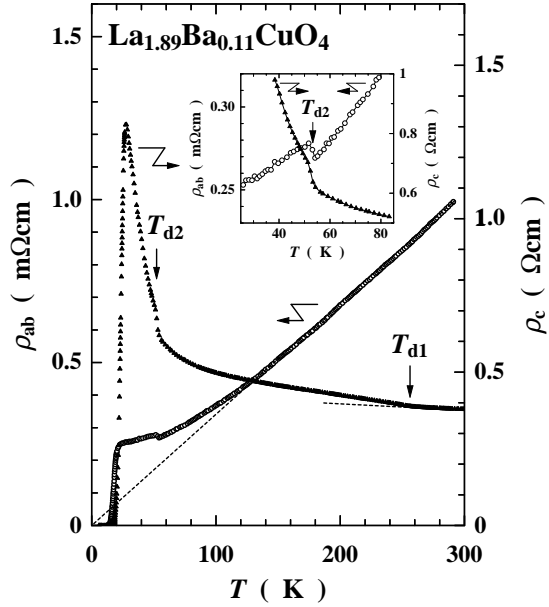}}
\end{center}
\caption{Temperature dependence of the in-plane ($\rho_{ab}$) and out-of-plane ($\rho_c$) electrical resistivities in the single-crystal La$_{2-x}$Ba$_x$CuO$_4$ with $x=0.11$. The $T_{d1}$ and $T_{d2}$ represent structural phase transition temperatures between the THT and OMT phases and between the OMT and TLT phases, respectively. The inset shows magnified plots of $\rho_{ab}$ and $\rho_c$ vs $T$ around $T_{d2}$.}
\label{fig:fig4} 
\end{figure}
\begin{figure}[htbp]
\begin{center}
\mbox{\epsfxsize=0.38 \textwidth \epsfbox{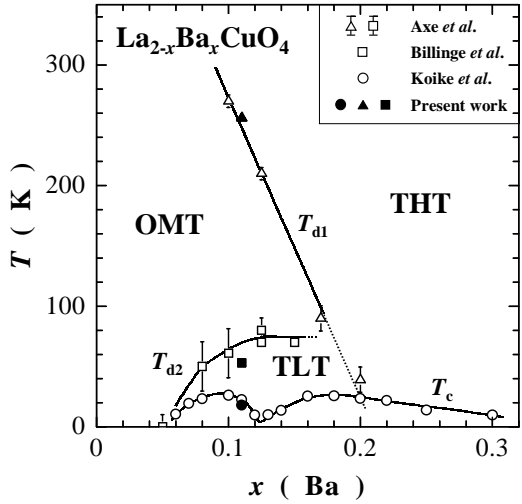}}
\end{center}
\caption{Phase diagram of La$_{2-x}$Ba$_x$CuO$_4$. Open circles represent $T_c$ from Ref. 41. Open triangles represent the structural phase transition temperature between the THT and OMT phases $T_{d1}$ from Ref. 10, and open squares represent that between the OMT and TLT phases $T_{d2}$ from Ref. 10 and 42. Filled circle, triangle and square represent the present $T_c$, $T_{d1}$ and $T_{d2}$ of the single-crystal La$_{2-x}$Ba$_x$CuO$_4$ with $x=0.11$, respectively.}
\label{fig:fig5} 
\end{figure}

The structural phase transition temperatures $T_{d1}$ and $T_{d2}$ of the single-crystal La$_{2-x}$Ba$_x$CuO$_4$ with $x=0.11$ have actually been estimated from the powder x-ray diffraction measurements at low temperatures. 
Figure 6(a) is a stack plot showing the temperature dependence of the diffraction profile of the (220)$_{THT}$ reflection in the notation of the THT phase, which is sensitive to the tetragonal-orthorhombic phase transitions. 
A single peak is observed at high temperatures above $T_{d1}$. 
The peak splits progressively with decreasing temperature below $T_{d1}$, corresponding to the (040)$_{OMT}$ and (400)$_{OMT}$ peaks in the notation of the OMT phase. 
Below $T_{d2}$, however, the two peaks merge into a broad single peak. 
To make clear the temperature dependence of the profile, the temperature dependence of the full width at half maximum, FWHM, of the (110)$_{THT}$ and (220)$_{THT}$ peaks are shown in Fig. 6(b) and 6(c), respectively. 
Below $T_{d1}$, the FWHM's increase gradually with decreasing temperature because of the second order transition at $T_{d1}$. 
Around $T_{d2}$, they decrease suddenly with decreasing temperature because of the first order transition at $T_{d2}$. 
The width of the transition at $T_{d2}$ is actually a little broad on account of the limited time of the measurements. 
Values of $T_{d1}$ and $T_{d2}$ are estimated to be $\sim$ 256 K and $\sim$ 53 K, respectively. 
Thus, it has been confirmed from the powder x-ray diffraction measurements that the kink and jump in the temperature dependence of the resistivity are indeed due to the structural phase transitions at $T_{d1}$ and $T_{d2}$, respectively. 
\begin{figure}[htbp]
\begin{center}
\mbox{\epsfxsize=0.47 \textwidth \epsfbox{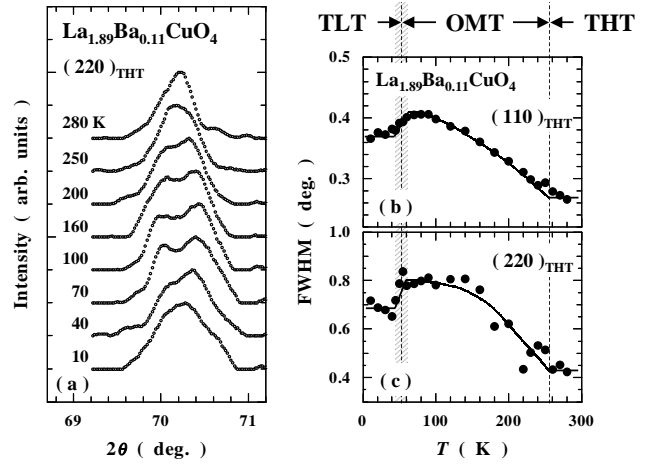}}
\end{center}
\caption{(a) Powder x-ray diffraction profiles of the (220)$_{THT}$ reflection in the notation of the THT phase due to Cu{\it K$_{\alpha 1}$} radiation at various temperatures for the single-crystal La$_{2-x}$Ba$_x$CuO$_4$ with $x=0.11$. (b), (c) Temperature dependence of the FWHM of the (110)$_{THT}$ and (220)$_{THT}$ peaks due to Cu{\it K$_{\alpha 1}$} radiation, respectively. Dashed lines at 256 K and 53 K represent $T_{d1}$ and $T_{d2}$ estimated from the resistivity measurements, respectively. Solid lines guide the reader's eye.}
\label{fig:fig6} 
\end{figure}

The temperature dependence of the in-plane thermoelectric power $S_{ab}$ and the in-plane Hall coefficient $R_H$ are shown in Fig. 7, together with the $\rho_{ab}$ vs $T$ plot. 
The $S_{ab}$ drops rapidly at $T_{d2}$ and decreases below $T_{d2}$ with decreasing temperature. 
Moreover, it changes the sign somewhat at low temperatures below $\sim$ 25 K, as shown in the inset of Fig. 7(b), which is well known to be characteristic of the 1/8 anomaly. \cite{sera,nakamura,koike4}
The $R_H$ also drops rapidly at $T_{d2}$ and decreases below $T_{d2}$ with decreasing temperature. 
Then, a conspicuous reversal in the sign of $R_H$ is observed at low temperatures below $\sim$ 25 K, where $S_{ab}$ also exhibits the sign reversal. 
These anomalous behaviors of $S_{ab}$ and $R_H$ are similar to those observed in La$_{1.6-x}$Nd$_{0.4}$Sr$_x$CuO$_4$ with $x=0.12$, though the sign reversal of $R_H$ is not well-defined in La$_{1.6-x}$Nd$_{0.4}$Sr$_x$CuO$_4$. \cite{nakamura}
\begin{figure}[htbp]
\begin{center}
\mbox{\epsfxsize=0.4 \textwidth \epsfbox{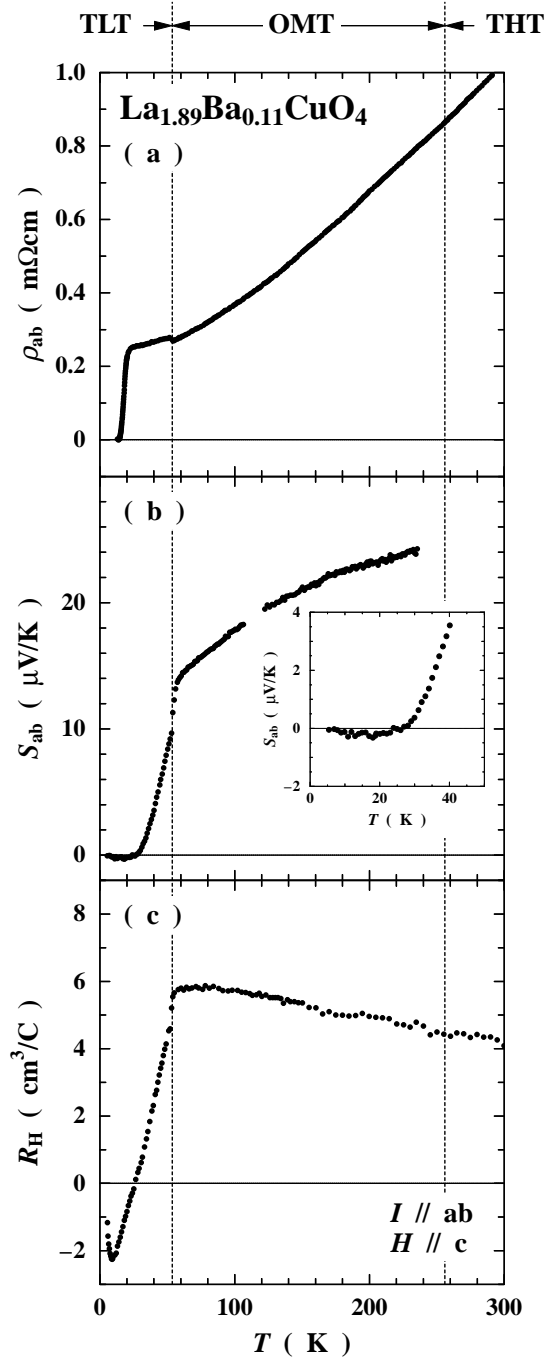}}
\end{center}
\caption{Temperature dependence of (a) the in-plane electrical resistivity $\rho_{ab}$, (b) the in-plane thermoelectric power $S_{ab}$ and (c) the in-plane Hall coefficient $R_H$ in the single-crystal La$_{2-x}$Ba$_x$CuO$_4$ with $x=0.11$. Dashed lines at 256 K and 53 K represent $T_{d1}$ and $T_{d2}$, respectively. The inset of (b) shows magnified plot of $S_{ab}$ below 50 K.}
\label{fig:fig7} 
\end{figure}
\section{Discussion}

We first discuss the clear jump or drop at $T_{d2}$ in the temperature dependence of $\rho_{ab}$, $\rho_c$, $S_{ab}$ and $R_H$, which we have observed for the first time in La$_{2-x}$Ba$_x$CuO$_4$. 
These clear transport anomalies at $T_{d2}$ have been formerly observed in La$_{1.6-x}$Nd$_{0.4}$Sr$_x$CuO$_4$ with $x\sim1/8$. \cite{nakamura}
However, this is a first observation of such anomalies in "La$_{2-x}$Ba$_x$CuO$_4$." \cite{first}
This is analogous to that observed in the polycrystalline La$_{2-x}$Ba$_x$CuO$_4$ with $x\sim1/8$. \cite{sera} 
Moreover, the anomalous behaviors are much clearer in the single crystal than in the polycrystal, as in the case of La$_{1.6-x}$Nd$_{0.4}$Sr$_x$CuO$_4$. \cite{nakamura,koike4}
These results demonstrate that carriers are affected by the structural phase transition at $T_{d2}$ in La$_{2-x}$Ba$_x$CuO$_4$ with $x=0.11$ as well as in La$_{1.6-x}$Nd$_{0.4}$Sr$_x$CuO$_4$ with $x=0.12$. \cite{nakamura}
That is, it is possible that the change of charge dynamics relevant to the change of lattice occurs at $T_{d2}$. 

In general, signs of the thermoelectric power and the Hall coefficient reflect the sign of carriers. 
Therefore, it is very likely that the observed sign reversals in the temperature dependences of $S_{ab}$ and $R_H$, as shown in Fig. 7, have the same origin. 
The sign reversal of $S_{ab}$ in the single-crystal La$_{2-x}$Ba$_x$CuO$_4$ with $x=0.11$ is very small, compared with that in the polycrystalline La$_{2-x}$Ba$_x$CuO$_4$ with $x=1/8$. \cite{sera}
It may be due to the smaller value of $x$ than 1/8 where the sign reversal is the most conspicuous in the polycrystalline La$_{2-x}$Ba$_x$CuO$_4$. 
As for the sign reversal of $R_H$, it is often observed in the superconducting fluctuation regime, though it has not yet been understood clearly. 
It is an empirical fact that the sign reversal of $R_H$ has a strong magnetic-field dependence in the superconducting fluctuation regime. 
That is, the magnitude of the sign reversal of $R_H$ decreases with increasing magnetic field, accompanied by the broadening of the superconducting transition curve in resistivity. \cite{matsuda} 
Figure 8 displays the temperature dependence of $R_H$ in various magnetic fields, measured using another batch of the single-crystal La$_{2-x}$Ba$_x$CuO$_4$ with $x=0.11$. 
It is found that the magnitude of the sign reversal of $R_H$ increases with increasing magnetic field and saturates in higher fields above 6 T. 
Moreover, the sign-reversal temperature is independent of the onset temperature of the superconducting transition, $T_c^{onset}$, estimated from the resistivity in each magnetic field. \cite{rho-h}
Therefore, it is difficult that the observed reversal in the sign of $R_H$ in the single-crystal La$_{2-x}$Ba$_x$CuO$_4$ with $x=0.11$ is attributed to the superconducting fluctuation. 
The origin of the sign reversals in $S_{ab}$ and $R_H$ is not clarified, but it is considered that these behaviors are characteristic of the 1/8 anomaly in a series of La-214 systems. \cite{sera,nakamura,koike4,koike6,adachi,geka}
\begin{figure}[htbp]
\begin{center}
\mbox{\epsfxsize=0.4 \textwidth \epsfbox{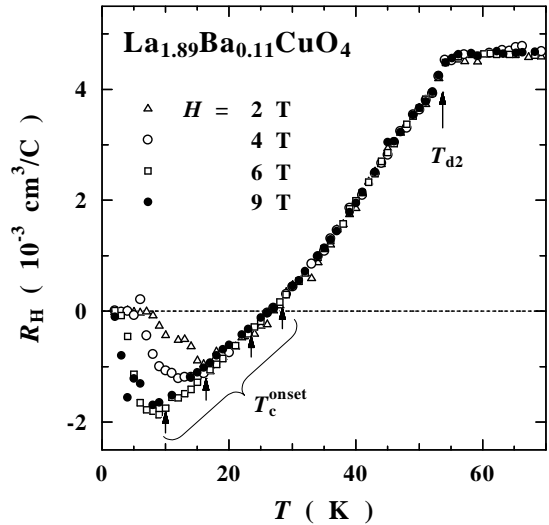}}
\end{center}
\caption{Temperature dependence of the in-plane Hall coefficient $R_H$ of the single-crystal La$_{2-x}$Ba$_x$CuO$_4$ with $x=0.11$ in various magnetic fields. The arrow at $\sim$ 53 K represents $T_{d2}$. The other arrows below $\sim$ 30 K represent the onset temperature of the superconducting transition, $T_c^{onset}$, in the respective magnetic fields.}
\label{fig:fig8} 
\end{figure}

We now discuss the experimental results in the single-crystal La$_{2-x}$Ba$_x$CuO$_4$ with $x=0.11$ from the view point of the static stripe order. 
The observed anomalous behaviors in $\rho_{ab}$, $\rho_c$, $S_{ab}$ and $R_H$ are quite similar to those observed in La$_{1.6-x}$Nd$_{0.4}$Sr$_x$CuO$_4$ with $x\sim1/8$ where the static stripe order is formed at low temperatures below $T_{d2}$. 
In particular, Noda {\it et al}. have suggested in La$_{1.4-x}$Nd$_{0.6}$Sr$_x$CuO$_4$ that the rapid decrease in $R_H$ below $T_{d2}$ is a typical behavior of the one-dimensional transport associated with the formation of the stripe order. \cite{noda}
Therefore, it is much convinced that the static stripe order of holes and spins is formed at low temperatures below $T_{d2}$ also in La$_{2-x}$Ba$_x$CuO$_4$ with $x=0.11$, though it has already been predicted from the experimental results in the polycrystalline samples. 
Although $\rho_{ab}$ exhibits an upturn at low temperatures below $T_{d2}$ in La$_{1.6-x}$Nd$_{0.4}$Sr$_x$CuO$_4$ with $x=0.12$, $\rho_{ab}$ in the single-crystal La$_{2-x}$Ba$_x$CuO$_4$ with $x=0.11$ does not exhibit any upturn but exhibits a metallic behavior. 
It is because the stripe order may be highly fluctuating at low temperatures below $T_{d2}$ in spite of the pinning by the TLT structure, as in the case of La$_{1.6-x}$Nd$_{0.4}$Sr$_x$CuO$_4$ with $x=0.10$. \cite{ichikawa} 
To make sure of the above speculation, a direct observation of the static stripe order from the neutron scattering experiment is now being planned. 

\section{Summary}

We have succeeded in growing high-quality single crystals of La$_{2-x}$Ba$_x$CuO$_4$ with $x=0.11$ by the TSFZ method under flowing O$_2$-gas of a high pressure (4 bars). 
We have investigated the temperature dependence of $\rho_{ab}$, $\rho_c$, $S_{ab}$ and $R_H$ and the crystal structure from the powder x-ray diffraction measurements at low temperatures. 
For the first time, we have observed a clear jump in both $\rho_{ab}$ and $\rho_c$ at $T_{d2}$. 
We have also found that both $S_{ab}$ and $R_H$ drop rapidly at $T_{d2}$ and decrease below $T_{d2}$ with decreasing temperature. 
These anomalous behaviors are analogous to, but much clearer than those observed in the polycrystalline La$_{2-x}$Ba$_x$CuO$_4$ with $x\sim1/8$. 
These results indicate that the change of charge dynamics relevant to the change of lattice occurs at $T_{d2}$ in La$_{2-x}$Ba$_x$CuO$_4$ with $x=0.11$. 
Moreover, these anomalous behaviors are quite similar to those observed in La$_{1.6-x}$Nd$_{0.4}$Sr$_x$CuO$_4$ with $x\sim1/8$ where the static stripe order appears at low temperatures below $T_{d2}$. 
It is much convinced that the static stripe order of holes and spins in the CuO$_2$ plane is formed below $T_{d2}$ in La$_{2-x}$Ba$_x$CuO$_4$ with $x\sim1/8$. 

\section*{Acknowledgments}

We are indebted to Prof. K. Takada and M. Ishikuro for their help in the ICP analysis. 
Valuable discussions with Dr. S. Watauchi and N. Ichikawa are gratefully acknowledged. 
This work was supported by a Grant-in-Aid for Scientific Research from the Ministry of Education, Science, Sports and Culture, Japan, and also by CREST of Japan Science and Technology Corporation. 
T. A. was supported by Research Fellowships of the Japan Society for the Promotion of Science for Young Scientists.


\begin{references}
\bibitem{la2o3}
It is well known that La$_2$O$_3$ changes to La(OH)$_3$ by absorbing water in air, leading to the destruction of the grown crystal.
\bibitem{tc}
The $T_c$ value estimated from the resistivity is higher than that from the susceptibility, which is also known in polycrystalline sample with $x\sim1/8$. \cite{mooden}
The phase of a little lower concentration of Ba than $x=0.11$ remaining in the filamentary shape may exhibit superconductivity at higher temperatures than 10.2 K.
\end{references}
\end{document}